\documentclass[11pt]{article}
\usepackage[letterpaper]{geometry}
\pdfoutput=1
\usepackage{amsmath,amsthm,amsfonts,amssymb}
\usepackage{enumerate,color,xcolor}
\usepackage{graphicx}
\usepackage{float}
\usepackage{enumitem}
\usepackage{breqn}
\usepackage{caption}
\usepackage{subcaption}

\usepackage{url}

\numberwithin{equation}{section}
\newcommand\numberthis{\addtocounter{equation}{1}\tag{\theequation}}
\theoremstyle{plain}

\newtheorem{theorem}{Theorem}

\newtheorem{proposition}[theorem]{Proposition}

\newtheorem{remark}[theorem]{Remark}

\newcommand{\add}[1]{\textcolor{red}{#1}}

\begin{document}

    \title{Diffusion Approximation of a Risk Model with Non-Stationary Hawkes Arrivals of Claims}
    \author{Zailei Cheng\thanks{Department of Mathematics, Florida State University, Tallahassee, FL 32306, USA;  \newline e-mail: zcheng@math.fsu.edu}
        \and
        Youngsoo Seol\thanks{Department of Mathematics, Dong-A
        	University, Busan, Saha-gu, Nakdong-daero 550, 37, Republic of
        	Korea;e-mail: prosul76@dau.ac.kr}}
    \maketitle

    \begin{abstract}
    We consider a classical risk process  with arrival of claims following a non-stationary Hawkes
    process. We study the asymptotic regime when the premium rate and the baseline intensity
    of the claims arrival process are large, and claim size is small.
    The main goal of the article is to establish a diffusion
    approximation by verifying a functional central limit theorem
    and to compute the ruin
    probability in finite-time horizon. Numerical results will also be given.
    \end{abstract}
    \maketitle

\section{Introduction}
In risk theory of insurance and finance literature, ruin is the most
important event. The theoretical foundation of ruin theory, known as the
Cram\'{e}r-Lundberg model or classical risk process was introduced by Lundberg~\cite{Lundberg}. In this paper we consider a classical risk process with the wealth at time $ t $ given by
\begin{equation}\label{riskprocess}
U_{t}=u+ct-\sum_{i=1}^{N_{t}}X_{i}.
\end{equation} where $X_{i}$ are i.i.d. claims with the first two moments
being finite, and independent of the claims arrival process $N_{t}$
which follows a non-stationary Hawkes process with the intensity \eqref{dynamics},
and $\rho>0$ is the constant premium rate that the insurance company receives,
and $u>0$ is the initial wealth of the insurance company.

In the classical risk model in~\cite{Lundberg}, $N_{t}$ is assumed to follow a
Poisson process, which has independent and stationary time increments. In
this paper, we assume that the arrival process $N_{t}$ follows a non-stationary Hawkes
process, which has the clustering and self-exciting features and the time
increments are dependent. A linear Hawkes process which was first introduced by A.G. Hawkes in 1971 \cite{Hawkes, Hawkes71II} is a simple point process $N$. In this paper, we consider the non-stationary Hawkes process. The stochastic intensity $\lambda$ of $ N $ at time $t$ is given by
\begin{equation}
\lambda({t}):= \mu + \int_{0}^{t-}h(t-s)N(ds) = \mu + \sum_{0<\tau_i<t}h(t-\tau_i),  \label{dynamics}
\end{equation}
where
$\tau_i$ are the occurrences of the points before time $t$, and 
$h(\cdot):[0, \infty) \rightarrow [0, \infty)$ and
we always assume that $\Vert h\Vert_{L^{1}}:=\int_{0}^{\infty}h(t)dt<\infty$.
We use the notation $N({t}):=N(0,t]$ to denote the number of
points in the interval $(0,t]$. When $h \equiv 0$, the non-stationary Hawkes process $N$ becomes a Poisson process with rate $\mu$. A commonly used nontrivial example of $h$ is an exponential function, i.e.,
$h(t)= \alpha e^{-\beta t}$ for $t \ge 0$, where $\alpha, \beta >0$. In this special case, the process $(\lambda, N)$
is Markovian. In the literature, the parameter $\mu$ is called the \textit{baseline intensity}, and $h(\cdot)$ is called
the \textit{exciting function} or sometimes referred to as the \textit{kernel function}. The linear Hawkes process exhibits both self--exciting (i.e., the occurrence
of an event increases the probabilities of future
events) and clustering properties. Hence it is very appealing in point process modeling and it has wide application
in various domains, including neuroscience \cite{Johnson96, Pernice2012, Reynaud2013}, seismology \cite{Ogata1988}, genome analysis \cite{Gusto2005, Reynaud2010}, social network \cite{Blundell2012, Crane2008}, finance (see the recent survey paper \cite{Bacry2015} and the references therein) and others.

A main topic in the mathematical finance or insurance literature, inspired by the
early contributions of Lundberg~\cite{Lundberg} and
Cram\'{e}r~\cite{cramer1930}, is the computation of the ruin
probability over both finite-time and infinite-time horizon. In fact, exact formulas for both finite-time and
infinite-time ruin probability are known only for few special
models. Therefore, asymptotic methods have been developed to derive
expansions of the ruin probability as the initial capital or reserve
increases to infinity. In this paper, we focus on computing the ruin probability over finite-time horizon.

Non-stationary Hawkes Process has wide application in insurance \cite{Stabile,ZhuRuin}. By applying the
techniques of large deviations, the asymptotics of the ruin
probabilities for risk processes in insurance were studied in
Stabile and Torrisi~\cite{Stabile} for the light-tailed claims and
in Zhu~\cite{ZhuRuin} for the heavy-tailed claims. However, these two papers focus on asymptotic regimes for large initial wealth. Our paper assumes baseline intensity for Hawkes Process, which can be used to study catastrophic events. Similar regime has been studied by Gao and Zhu \cite{GZ2}, where they use large initial intensity for Markovian case. Our paper studies different asymptotic regime that is when the baseline intensity of arrival process is large. We apply functional central limit theorem to obtain approximations and use that to study finite time ruin probability. The limit
theorems have also been studied for an extension of linear Hawkes
processes and Cox-Ingersoll-Ross processes in Zhu~\cite{Zhucir},
which has applications in short interest rate models in finance.

A diffusion approximation is constructed for an insurance risk model
which was considered by Embrechts and Schmidli~\cite{Embrechts},
where the company is allowed to borrow money if needed and to invest
money for large surpluses. Moreover, diffusion approximations of the
risk reserve process were first studied by Iglehart~\cite{Iglehart}
and subsequently by Grandell~\cite{Grandell},
Harrison~\cite{Harrison}, Schmidli~\cite{Schmidli},  and
Bauerle~\cite{Bauerle} by using the machinery of weak convergence.

The main goal of this article is to develop diffusion approximations for the wealth process $U_{t}$ which was introduced in~\eqref{riskprocess} under
the regime when the premium rate and the baseline intensity of the
claims arrival process are large, and claim size is small. Furthermore, employing approximations of risk processes, we obtain formulas for ruin probabilities in finite horizon. Finally, we give the numerical illustrations for the results.

The rest of the paper is organized as the follows. In Section 2, we
state the main results on the functional central limit theorem for
aggregate claims process and hence also the wealth process, where the
claims arrive according to a non-stationary Hawkes process. In Section
3, we obtain the finite-horizon ruin
probability asymptotic for diffusion approximation with large
initial wealth. Finally, in Section 4, we give some examples for
numerical results. The proofs of the main result are given in the
Appendix.
\section{Functional Central Limit Theorem
        for Aggregate Claims Process}
    In this section, we study
    approximations for the aggregate claims process with a large baseline
    intensity.
More precisely, we consider
\begin{equation}
U_{t}^{\mu}=u+\rho^{\mu}t-\sum_{i=1}^{N_{t}^{\mu}}\frac{1}{\sqrt{\mu}}Y_{i},
\end{equation}
so that the claim sizes are scaled by a factor $\frac{1}{\sqrt{\mu}}$ and $Y_{i}$ are i.i.d.
with first two moments being finite, and we define $\mathbb{E}[Y_{1}]=m_{1}$, $\mathbb{E}[Y_{1}^{2}]=m_{2}$.,
and $\rho^{\mu}=\sqrt{\mu}\frac{t}{1-\Vert h\Vert_{L^{1}}}m_{1}+c$ for some constant $c>0$.
We assume the claims arrival process $N^{\mu}$ has intensity given by \eqref{dynamics}.
We write $N^{\mu}$
to emphasize that the baseline intensity of this Hawkes process is $\mu$. Our goal is to
establish a functional central limit theorem for the $U_{t}^{\mu}$ process
in the asymptotic regime $\mu \rightarrow \infty$.

In the classical risk model when the claims arrival process follows a standard Poisson process
with constant intensity $\mu$, this is the standard diffusion approximation that is used
in the insurance literature.

\begin{proposition}\label{prop1}
	
	We first present the mean and variance of the
	arrival process $N_{t}^{\mu}$.
	
    \begin{enumerate}[label=(\alph*)]
        \item $ \mathbb{E}[N_t^{\mu}]=\mu\int_{0}^{t}g_1(s)ds $,
        \item $ \mbox{Var}[N_t^{\mu}]=\mu\int_{0}^{t}g_2(s)ds $,
    \end{enumerate}

	where
\begin{equation} \label{eq:var}
g_2(t)=\int_{0}^{t}h(t-s)g_2(s)ds+g_1(t)^2,
\end{equation}
and $g_1(t)$ satisfies the integral equation:
\begin{equation} \label{eq:phi}
g_1(t)=1+\int_{0}^{t}h(t-s)g_1(s)ds.
\end{equation}
\end{proposition}

We now present a result on the functional central limit theorem (FCLT) for the aggregate claims process,
and hence also the wealth process,
where the claims arrive according to a non-stationary Hawkes process.
Write $(D([0,\infty),\mathbb{R}),J_{1})$ as the space of c\`{a}dl\`{a}g processes on $[0, \infty)$ that are equipped with Skorohod $J_1$
topology (see, e.g., Billingsley \cite{Billingsley}).

Let us denote the aggregate claims process as:
\begin{equation}
X_{t}^{\mu}=\sum_{i=1}^{N_{t}^{\mu}}Y_{i}.
\end{equation}

\begin{theorem} \label{thm:FCLT}
    Assume that $h(\cdot)$ is a decreasing function and $\int_0^\infty t \cdot h(t) dt<\infty$. As $\mu\rightarrow\infty$,
    \begin{equation*}
    \frac{X^{\mu}_{t}-\mu m_1\int_{0}^{t}g_1(s)ds}{\sqrt{\mu}}\Rightarrow G,
    \end{equation*}
    weakly in $(D([0,\infty),\mathbb{R}),J_{1})$, where $G$ is a mean-zero almost surely continuous Gaussian process
    with the covariance function, $t \ge s$,
    \begin{align}\label{eq:cov-G}
    \mbox{Cov}(G(t),G(s))
    =m_1^2\mbox{Cov}(N_t^1,N_s^1)
    +\mathbb{E}[N_s^1](m_2-m_1^2).
    \end{align}
    As a result,
    \begin{equation}
    U_{t}^{\mu}\rightarrow u+ct-G(t),
    \end{equation}
    weakly in $(D([0,\infty),\mathbb{R}),J_{1})$.
\end{theorem}

The key in the observation is that $G(t)$ can be written as an
integral of a centered Gaussian process plus an
independent Brownian motion:

\begin{proposition}
    It holds in distribution that
    \begin{equation}
    G(t)=m_{1}\int_{0}^{t}H(s)ds+\frac{m_{2}}{1-\Vert h\Vert_{L^{1}}}B_{t},
    \end{equation}
    where $H(s)$ is a centered Gaussian process
    with for any $t\geq s$,
    \begin{equation}
    \mbox{Cov}(H(t),H(s))=\mbox{Cov}(N_t^1,N_s^1),
    \end{equation}
    and $B_{t}$ is a standard Brownian motion independent of $H(t)$.
\end{proposition}

\begin{remark}
   We first briefly explain why we obtain a Gaussian limit $G(t)$. By the immigration--birth representation of Hawkes processes (see, e.g., \cite{Hawkes71II}),
   we know that for a Hawkes process $N^{\mu}$ with a baseline intensity $\mu$ and an exciting function $h$, we can decompose it as the sum of $\mu$ independent Hawkes process, each having a baseline intensity one and an exciting function $h$. Then one expects by central limit theorem type of arguments, $N^{\mu}$ will be asymptotically Gaussian when we send $\mu$ to infinity.
\end{remark}

\begin{remark}
    We next discuss the variance function of $G$ in \eqref{eq:cov-G}.
    In general, the variance function of $G$ in \eqref{eq:cov-G} is semi-explicit and we can compute it by first numerically solving $g_1$ and $ g_2 $ via the integral equation \eqref{eq:var} and \eqref{eq:phi}.
    In the special case when $h(t)= \alpha e^{-\beta t}$ where $\alpha< \beta$, the variance function of $G$ is explicit. To see this, we first deduce from \eqref{eq:var} and \eqref{eq:phi} that

    \begin{equation}
    g_1(t)=\frac{\alpha}{\alpha-\beta}e^{(\alpha-\beta)t}-\frac{\beta}{\alpha-\beta}
    \end{equation}and
    \begin{equation}
    g_2(t)=\frac{2\alpha^2}{(\alpha-\beta)^2}e^{2(\alpha-\beta)t}-\left[\frac{\alpha(\alpha+\beta)}{(\alpha-\beta)^2}+\frac{2\alpha\beta t}{\alpha-\beta}\right]e^{(\alpha-\beta)t}+\frac{\beta}{\beta-\alpha}.
    \end{equation}

    Then from Proposition 1, we get

    \begin{equation}\label{expectofn}
    \mathbb{E}[N_t^1]=\frac{\alpha}{(\alpha-\beta)^2}\left[e^{(\alpha-\beta)t}-1\right]-\frac{\beta t}{\alpha-\beta}
    \end{equation} and
    \begin{equation}\label{varofn}
    \mbox{Var}[N_t^1]=\frac{\alpha^2}{(\alpha-\beta)^3}\left[e^{2(\alpha-\beta)t}-1\right]-\frac{\alpha}{(\alpha-\beta)^2}\left[e^{(\alpha-\beta)t}-1\right]-\frac{2\alpha\beta t}{(\alpha-\beta)^2}e^{(\alpha-\beta)t}+\frac{\beta t}{\beta-\alpha}.
    \end{equation}

    Let $ s=t $ in \eqref{eq:cov-G}, we get:

    \begin{align}\label{eq:Kt-exp}
    &\mbox{Var}(G(t))=m_1^2\mbox{Var}[N_t^1]+\mathbb{E}[N_t^1](m_2-m_1^2)\nonumber\\
    &=m_1^2\bigg(\frac{\alpha^2}{(\alpha-\beta)^3}\left[e^{2(\alpha-\beta)t}-1\right]-\frac{\alpha}{(\alpha-\beta)^2}\left[e^{(\alpha-\beta)t}-1\right]-\frac{2\alpha\beta t}{(\alpha-\beta)^2}e^{(\alpha-\beta)t}+\frac{\beta t}{\beta-\alpha}\bigg)\nonumber\\
    &+(m_2-m_1^2)\bigg(\frac{\alpha}{(\alpha-\beta)^2}\left[e^{(\alpha-\beta)t}-1\right]-\frac{\beta t}{\alpha-\beta}\bigg).
    \end{align}

\end{remark}

\begin{remark} \label{rem6}
Furthermore we are able to calculate the covariance of $ G $. Please see the derivation in appendix. For $t \ge s$,

\begin{align*}\label{eq:cov-G-exponential}
&\mbox{Cov}(G(t),G(s))\\
&=m_1^2\bigg(\mbox{Var}[N_s^1]+(\mathbb{E}[N_s^1])^2+\mathbb{E}[N_s^{1}(N_t^{1}-N_s^{1})]-\mathbb{E}[N_s^{1}]\mathbb{E}[N_t^{1}]\bigg)+\mathbb{E}[N_s^1](m_2-m_1^2)\\
&=m_1^2\bigg(\frac{\alpha^2}{(\alpha-\beta)^3}\left[e^{2(\alpha-\beta)t}-1\right]-\frac{\alpha}{(\alpha-\beta)^2}\left[e^{(\alpha-\beta)t}-1\right]-\frac{2\alpha\beta t}{(\alpha-\beta)^2}e^{(\alpha-\beta)t}+\frac{\beta t}{\beta-\alpha}\\
&+\left[\frac{\alpha}{(\alpha-\beta)^2}\left[e^{(\alpha-\beta)s}-1\right]-\frac{\beta s}{\alpha-\beta}\right]^2\\
&+\left[M_1e^{(\alpha-\beta)s}+\frac{\alpha}{(\alpha-\beta)^2}\left[e^{(\alpha-\beta)s}-1\right]-\frac{\beta s}{\alpha-\beta}\right](t-s)\\
&+M_2\left[e^{(\beta-\alpha)t}-e^{(\beta-\alpha)s}\right]+M_3\left[e^{(\alpha-\beta)t}-e^{(\alpha-\beta)s}\right]+M_4\left[e^{2(\alpha-\beta)t}-e^{2(\alpha-\beta)s}\right]\\
&-\left[\frac{\alpha}{(\alpha-\beta)^2}\left[e^{(\alpha-\beta)s}-1\right]-\frac{\beta s}{\alpha-\beta}\right]\left[\frac{\alpha}{(\alpha-\beta)^2}\left[e^{(\alpha-\beta)t}-1\right]-\frac{\beta t}{\alpha-\beta}\right]\bigg)\\
&+(m_2-m_1^2)\bigg(\frac{\alpha}{(\alpha-\beta)^2}\left[e^{(\alpha-\beta)s}-1\right]-\frac{\beta s}{\alpha-\beta}\bigg), \numberthis
\end{align*}

where $ M_i, i\in 1,2,3,4 $ are constants and

\begin{equation*}
\begin{split}
&M_1=\frac{(2\beta-\alpha)[-2\alpha^2(2+\alpha)+\alpha^2(2+\beta)+\alpha^2(2+2\alpha-\beta)-2\alpha]}{2\alpha(\beta-\alpha)^3}\\
&M_2=\frac{\alpha(2+\beta)}{2(\beta-\alpha)^2}\\
&M_3=\frac{\alpha(2+2\alpha-\beta)(2\beta-\alpha-\alpha\beta)-(\beta-\alpha)(2\alpha+2)}{2(\beta-\alpha)^3}\\
&M_4=\frac{-\alpha(3\beta+\alpha\beta-2\alpha-\alpha^2)}{(\beta-\alpha)^3}.
\end{split}
\end{equation*}

In this special case, we notice that the variance function of $G$, is nonlinear in $t$ in general. This is very different from the case when $N^{\mu}$ is a Poisson process (i.e., $h \equiv 0$) where $G$ becomes a standard Brownian motion.
\end{remark}

\section{Ruin Probability for the Diffusion Approximation}
In this section, we focus on developing the asymptotic estimates for the finite-horizon ruin probabilities.
 In the large baseline intensity limit, the ruin probability becomes:
\begin{equation}\label{ruin2}
\mathbb{P}\left(\sup_{0\leq t\leq T}\{G(t)-ct\}>u\right),
\end{equation}
for the finite-horizon case.

From the fact that \begin{equation}
    U_{t}^{\mu}\rightarrow u+ct-G(t),
    \end{equation}
    weakly in $(D([0,\infty),\mathbb{R}),J_{1})$ in Theorem~\ref{thm:FCLT}, it suffices to study \eqref{ruin2} as the large baseline intensity approximation to get the finite-horizon ruin probabilities for $U_{t}^{\mu}$.


Next, let us consider the exact asymptotics for the finite-time ruin probability
with large initial wealth. We rely on the results in \cite{Debickietal}.

Let $\sigma(t)$ be the standard deviation function
of $G(t)$. Let us consider $t\in[0,T]$.
We know that
\begin{equation}
\sigma(t)=\sqrt{m_1^2\int_{0}^{t}g_2(s)ds+(m_2-m_1^2)\int_{0}^{t}g_1(s)ds},
\end{equation}
which is increasing in $t$ with unique maximum achieved at $t=T$.
We can compute that
\begin{equation}
\sigma(t)
=\sigma(T)-\frac{1}{2\sigma(T)}
\left(m_1^2g_2(T)+(m_2-m_1^2)g_1(T)\right)(T-t)+o(T-t),
\end{equation}
as $t\rightarrow T$.
For any $t>s$,
\begin{equation}
\mbox{Cov}(G(t),G(s))=m_1^2\left[\mathbb{E}[(N_s^{1})^2]+\mathbb{E}[N_s^{1}(N_t^{1}-N_s^{1})]-\mathbb{E}[N_s^{1}]\mathbb{E}[N_t^{1}]\right]+\mathbb{E}[N_s^1](m_2-m_1^2).
\end{equation}
We can further compute that
\begin{align} \label{covggsi}
&\mbox{Cov}\left(\frac{G(t)}{\sigma(t)},\frac{G(s)}{\sigma(s)}\right)
\\ \nonumber
&=1+\left(\frac{m_1^2(2\mbox{Cov}(N_T^1,\lambda_T)-g_2(T))-(m_2-m_1^2)g_1(T)}{2\sigma(T)^2}\right)(t-s)+o(t-s)
\end{align}
as $t>s$, $ t-s\rightarrow 0 $ and $s\rightarrow T$.

The Assumption A1 is thus satisfied in \cite{Debickietal} for $G(t)$.
The Assumption A2 trivially holds in \cite{Debickietal} for $G(t)$, see proof of Theorem~\ref{thm:FCLT} in appendix.

By Theorem 3.1. \cite{Debickietal}, we have
\begin{equation} \label{P2}
\mathbb{P}\left(\sup_{0\leq t\leq T}\{G(t)-ct\}>u\right)
\sim
\mathcal{P}_{1}^{\tilde{N}/\tilde{G}}
\Psi\left(\frac{u+cT}{\sigma(T)}\right),
\end{equation}
as $u\rightarrow\infty$, where
$\tilde{N}=\frac{1}{2}
\left(m_1^2g_2(T)+(m_2-m_1^2)g_1(T)\right)$ and \\ $\tilde{G}=\frac{m_1^2(2\mbox{Cov}(N_T^1,\lambda_T)-g_2(T))-(m_2-m_1^2)g_1(T)}{2}$,
where
\begin{equation}
\mathcal{P}^{R}_{\alpha}:=\lim_{S\rightarrow\infty}
\mathbb{E}\left[\exp\left(\sup_{0\leq t\leq S}\{\sqrt{2}B_{\alpha/2}(t)-(1+R)t^{\alpha}\}\right)\right],
\end{equation}
where $B_{\alpha/2}$ is a fractional Brownian motion with Hurst index $\alpha/2$ and $0<\alpha\leq 2$ and $R>0$.

In our setting, $\alpha=1$, and $B_{1/2}$ is a standard Brownian motion.

\section{Numerical Studies}
In this section we study several numerical illustrations for
    the theoretical results of this article.
For $h(t)=\alpha e^{-\beta t}$, we
can simulate the Gaussian process $G(t)$, and numerically compute
the ruin probability for the finite-horizon case.

\subsection{Jump size as exponential distribution}

Here suppose $Y_i$ follows an i.i.d. exponential distribution with intensity $\lambda$, i.e. $p(x)=\lambda e^{-\lambda x}$, we computed the ruin probability with different parameters $\alpha,\beta,T,\lambda,c,u.$ We recall that in Remark~\ref{rem6}, we assume $ \alpha<\beta $ so that the covariance function of $ G $ is explicit.

\begin{figure}[H]
    \centering
    \begin{subfigure}[b]{0.47\textwidth}
        \includegraphics[width=\linewidth]{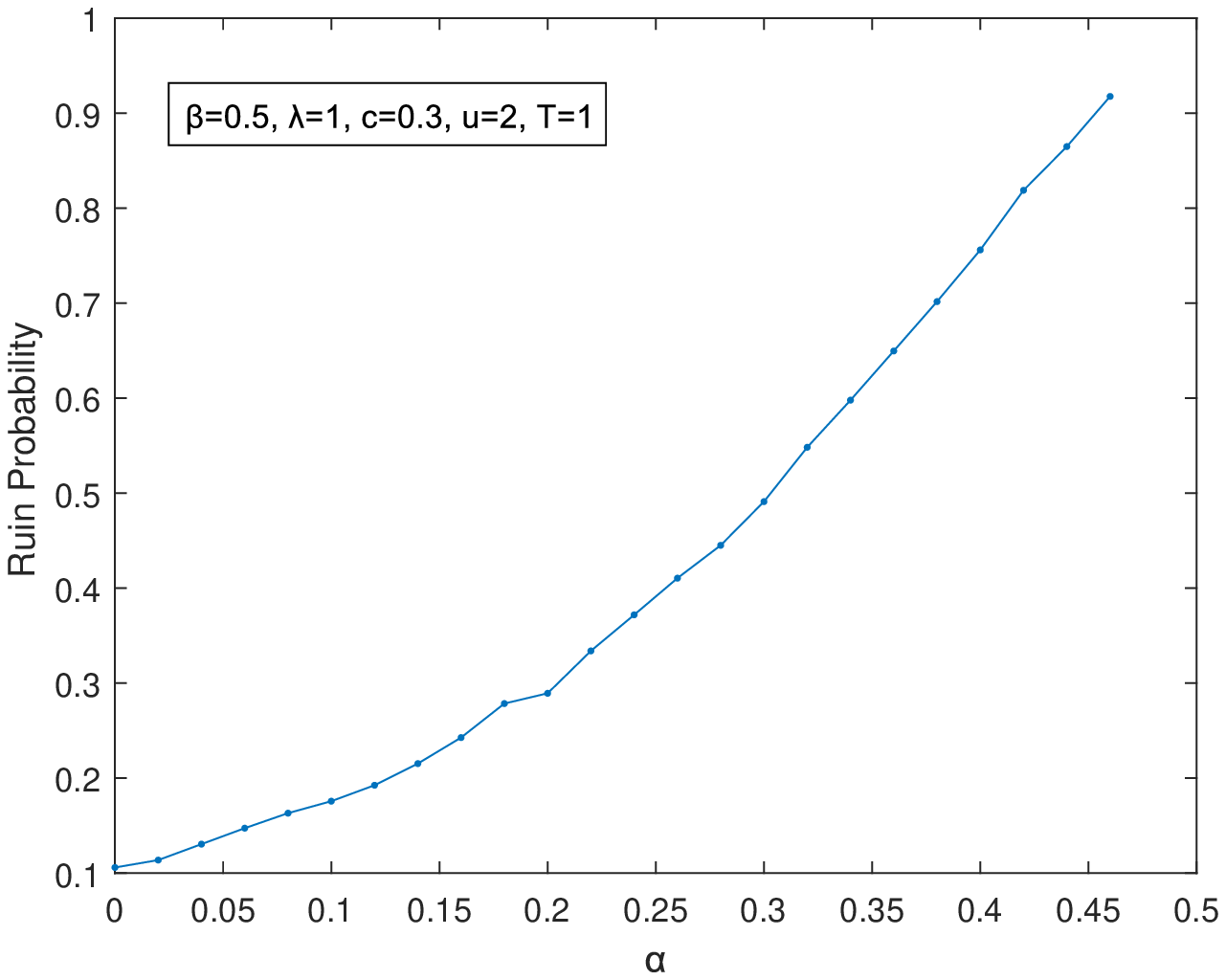}
        \caption{Ruin probability as $\alpha$ increases. (Here we take $\beta=0.5, \lambda=1, c=0.3, u=2, T=1$.)}
        \label{fig:alpha}
    \end{subfigure}
    ~
    \begin{subfigure}[b]{0.47\textwidth}
        \includegraphics[width=\linewidth]{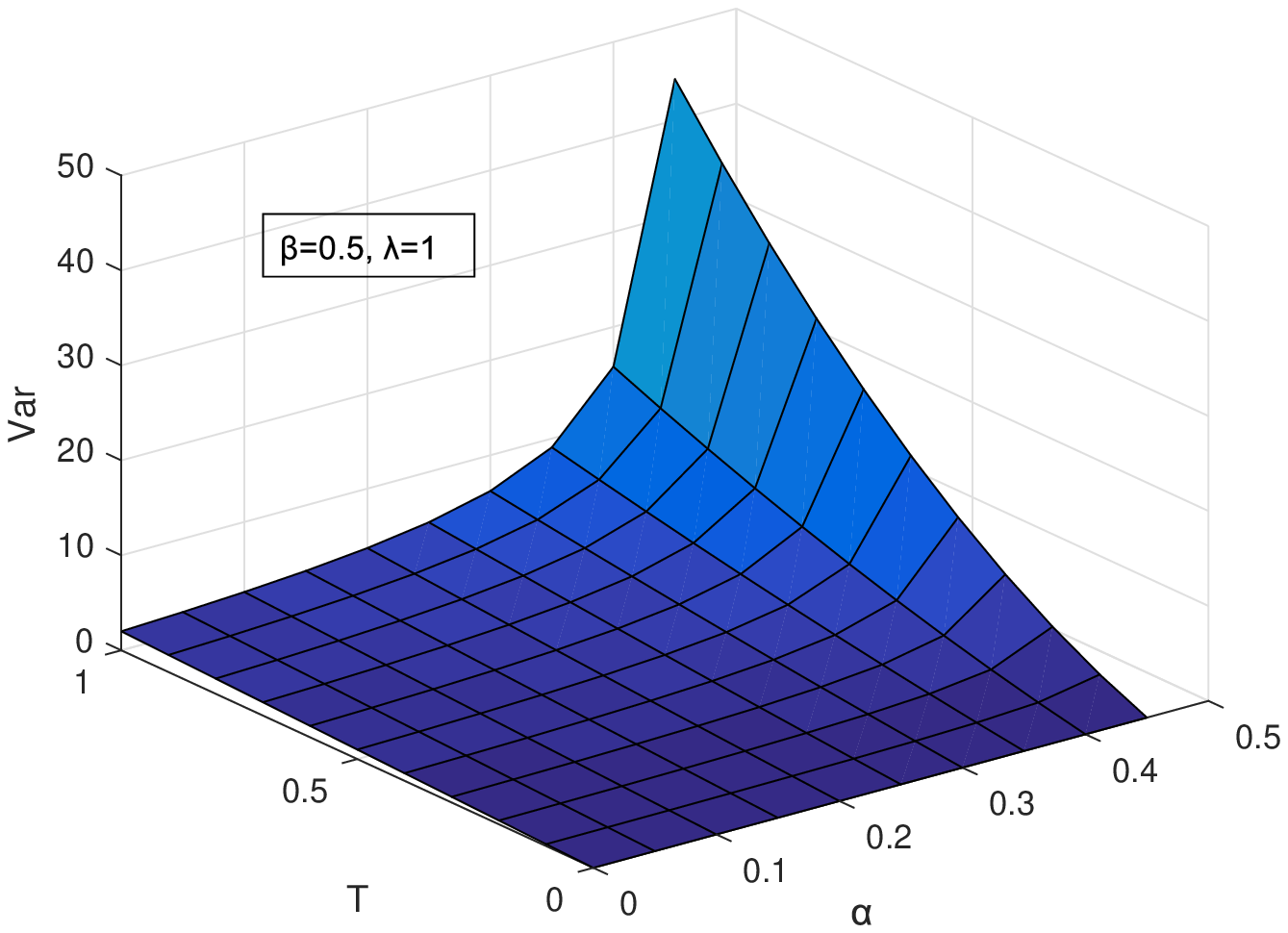}
        \caption{Variance of $G(t)$ as a function of $\alpha$ and $T$. (Here we take $\beta=0.5, \lambda=1$)}
        \label{fig:alphavar}
    \end{subfigure}
    \caption{Ruin Probability with varying $ \alpha $.}
\end{figure}

From Figure 1(a) we can see that with other parameters fixed, the ruin probability is an increasing function of $\alpha$. To explain, we plotted the variance of $G(t)(0\leq t\leq T)$ shown in \eqref{eq:Kt-exp}, as a function of $\alpha$ and $T$. In Figure 1(b), the variance increases as $\alpha$ increases. Intuitively, as the variance of $G(t)$ increases, the probability that $G(t)$ exceed a certain range increases. So the ruin probability increases.

Then we plot the ruin probability as a function of the intensity of the jump $\lambda$. And also, we can infer this from the plot of $\text{Var}(G(t))$ versus $\lambda$:

\begin{figure}[H]
    \centering
    \begin{subfigure}[b]{0.47\textwidth}
        \includegraphics[width=\linewidth]{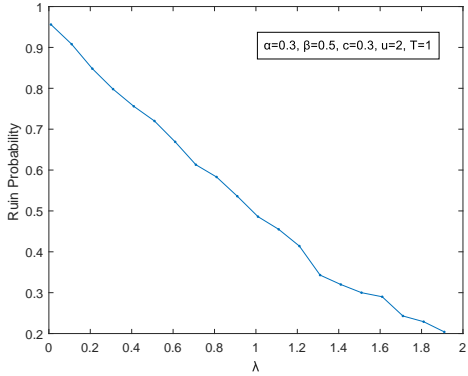}
        \caption{Ruin probability as $\lambda$ increases. (Here we take $\alpha=0.3, \beta=0.5, c=0.3, u=2$, $T$=1.)}
        \label{fig:lambda}
    \end{subfigure}
    ~
    \begin{subfigure}[b]{0.47\textwidth}
        \includegraphics[width=\linewidth]{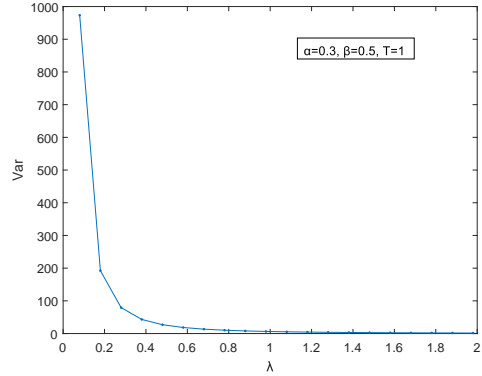}
        \caption{Variance of $G(t)$ as a function of $\lambda$. (Here we take $\alpha=0.3, \beta=0.5$.)}
        \label{fig:lambdavar}
    \end{subfigure}
    \caption{Ruin Probability with varying $ \lambda $.}
\end{figure}

\subsection{Jump size as Gamma distribution}

Assume the jump size $Y_i$ follows Gamma distribution with shape $a$ and rate $b$, i.e. $p(x;a,b)=\frac{b^ax^{a-1}e^{-bx}}{\Gamma(a)}$. Let's see how the ruin probability changes as $a$ change.

\begin{figure}[H]
    \centering
    \begin{subfigure}[b]{0.47\textwidth}
        \includegraphics[width=\linewidth]{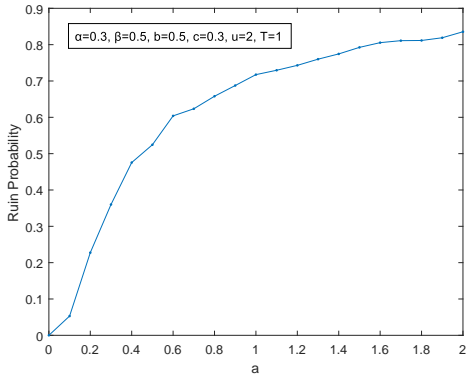}
        \caption{Ruin probability as $a$ increases. (Here we take $\alpha=0.3, \beta=0.5, b=0.5, c=0.3,u=2$, $T$=1.)}
        \label{fig:a}
    \end{subfigure}
    ~
    \begin{subfigure}[b]{0.47\textwidth}
        \centering
        \includegraphics[width=\linewidth]{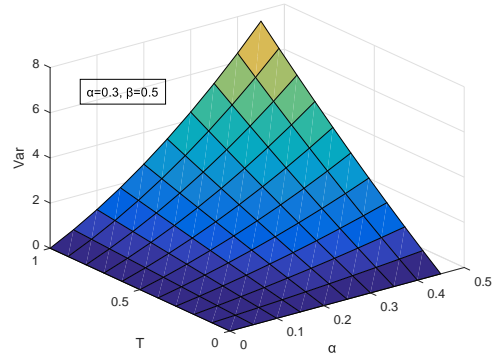}
        \caption{Variance of $G(t)$ as a function of $a$ and $T$. (Here we take $\alpha=0.3, \beta=0.5$.)}
        \label{fig:avar}
    \end{subfigure}
    \caption{Ruin Probability with varying $ a $.}
\end{figure}

We can see that the ruin probability is an increasing function of $a$, and this can be explained by the $\text{Var}(G(t))$ in Figure 3(b), the variance increases as $a$ increases.

\section{Appendix}

\subsection{Proof of Proposition~\ref{prop1}}

\begin{proof}
By the result of the moment generating function of $ N_t^{\mu} $ obtained in Zhu~\cite{ZhuCLT1}, we have, for any $ \theta\in\mathbb{R} $ and $ \theta<\Vert h\Vert_{L^{1}}-1-\log\Vert h\Vert_{L^{1}} $,
\begin{equation}\label{EN}
\mathbb{E}[e^{\theta N_t^{\mu}}]=e^{\mu\int_{0}^{t}(F_N(s)-1)ds},
\end{equation}
where the function $ F_N $ is the unique solution to the integral equation
\begin{equation}\label{FNT}
F_N(t)=e^{\theta}\mathbb{E}[e^{\int_{0}^{t}h(s)(F_N(s)-1)ds}].
\end{equation}
We first compute the first two moments of $ N_t^{\mu} $. By differentiating the moment generating function of $ N_t^{\mu} $ with respect to $ \theta $ in \eqref{EN}, we get
\begin{equation}
\frac{\partial}{\partial\theta}\mathbb{E}[e^{\theta N_t^{\mu}}]=\mu\int_{0}^{t}\frac{\partial}{\partial\theta}F_N(s)dse^{\mu\int_{0}^{t}(F_N(s)-1)ds},
\end{equation}
and by differentiating with respect to $ \theta $ again, we get
\begin{equation}
\frac{\partial^2}{\partial\theta^2}\mathbb{E}[e^{\theta N_t^{\mu}}]=\mu\int_{0}^{t}\frac{\partial^2}{\partial\theta^2}F_N(s)dse^{\mu\int_{0}^{t}(F_N(s)-1)ds}+\left(\mu\int_{0}^{t}\frac{\partial}{\partial\theta}F_N(s)ds\right)^2e^{\mu\int_{0}^{t}(F_N(s)-1)ds}.
\end{equation}
By differentiating both sides of \eqref{FNT} w.r.t. $ \theta $, we get
\begin{equation}\label{partialF}
\frac{\partial}{\partial\theta}F_N(t)=\mathbb{E}\left[\left(1+\int_{0}^{t}h(s)\frac{\partial}{\partial\theta}F_N(t-s)ds\right)e^{\theta+\int_{0}^{t}h(s)(F_N(t-s)-1)ds}\right].
\end{equation}

By differentiating again w.r.t. $ \theta $, we get

\begin{align}\label{ppartialF}
\frac{\partial^2}{\partial\theta^2}F_N(t)=\mathbb{E}\left[\left(1+\int_{0}^{t}h(s)\frac{\partial}{\partial\theta}F_N(t-s)ds\right)^2e^{\theta+\int_{0}^{t}h(s)(F_N(t-s)-1)ds}\right]\\ \nonumber
+\mathbb{E}\left[\int_{0}^{t}h(s)\frac{\partial^2}{\partial\theta^2}F_N(t-s)dse^{\theta+\int_{0}^{t}h(s)(F_N(t-s)-1)ds}\right].
\end{align}
By letting $ \theta=0 $ in \eqref{partialF}, we get
\begin{equation*}
\frac{\partial}{\partial\theta}F_N(t)\bigg|_{\theta=0}=1+\int_{0}^{t}h(s)\frac{\partial}{\partial\theta}F_N(t-s)\bigg|_{\theta=0}ds.
\end{equation*}
This implies that
\begin{equation*}
\frac{\partial}{\partial\theta}F_N(t)\bigg|_{\theta=0}=g_1(t),
\end{equation*}
where $ g_1 $ is defined in \eqref{eq:phi} and thus
\begin{equation*}
\mathbb{E}[N_t^{\mu}]=\frac{\partial}{\partial\theta}\mathbb{E}[e^{\theta N_t^{\mu}}]\bigg|_{\theta=0}=\mu\int_{0}^{t}\frac{\partial}{\partial\theta}F_N(s)\bigg|_{\theta=0}ds=\mu\int_{0}^{t}g_1(s)ds.
\end{equation*}
By letting $ \theta=0 $ in \eqref{ppartialF}, we get
\begin{equation*}
\frac{\partial^2}{\partial\theta^2}F_N(t)\bigg|_{\theta=0}=(g_1(t))^2+\int_{0}^{t}h(s)\frac{\partial^2}{\partial\theta^2}F_N(t-s)\bigg|_{\theta=0}ds.
\end{equation*}
By the definition of $ g_2 $ in \eqref{eq:var}, we have $ \frac{\partial^2}{\partial\theta^2}F_N(t)\bigg|_{\theta=0}=g_2(t) $. Finally we conclude that
\begin{align*}
\mathbb{E}[(N_t^{\mu})^2]
&=\frac{\partial^2}{\partial\theta^2}\mathbb{E}[e^{\theta N_t^{\mu}}]\bigg|_{\theta=0}\\
&=\mu\int_{0}^{t}\frac{\partial^2}{\partial\theta^2}F_N(s)\bigg|_{\theta=0}ds+\left(\mu\int_{0}^{t}\frac{\partial}{\partial\theta}F_N(s)\bigg|_{\theta=0}ds\right)^2\\
&=\mu\int_{0}^{t}g_2(s)ds+\left(\mu\int_{0}^{t}g_1(s)ds\right)^2.
\end{align*}
So we have
\begin{equation*}
\mbox{Var}[N_t^{\mu}]=\mathbb{E}[(N_t^{\mu})^2]-(\mathbb{E}[N_t^{\mu}])^2=\mu\int_{0}^{t}g_2(s)ds.
\end{equation*}
\end{proof}

\subsection{Proof of Theorem~\ref{thm:FCLT}}

\begin{proof}
    For the sake of simplicity, we
assume that $\mu\in\mathbb{N}.$ The argument to go from $\mu\in\mathbb{N}$ to
non-integer-valued $\mu$ follows the same argument as in Gao and Zhu~\cite{GZ3} . By
    immigration birth representation, we can decompose $N^{\mu}$ as the
    sum of $\mu$ independent and identically distributed (i.i.d) Hawkes
    processes $N_i^1, i =1, 2,\ldots, \mu$, each distributed as a
     Hawkes process with base intensity $1$ (the superscript 1
    in $N_i^1$) and the exciting function $h(\cdot)$. For notational
    simplicity, we use $N_i(\cdot)$ for $N_i^1 (\cdot)$. As a result, we
    can decompose $X^{\mu}$ as the sum of $\mu$ i.i.d. compound Hawkes
    processes and let us write $X^{\mu}=\sum_{i=1}^{\mu}X^{i}_{t}$.

    Therefore,
    \begin{equation*}
    \frac{X_t^{\mu}-\mu m_1\int_{0}^{t}g_1(s)ds}{\sqrt{\mu}}
    =\frac{1}{\sqrt{\mu}}\sum_{i=1}^{\mu}\left[X^{i}_{t}-m_1\int_{0}^{t}g_1(s)ds\right].
    \end{equation*}

    Let $\tilde{N}_t^i:=N_t^i-\int_{0}^{t}g_1(s)ds$.
    Then, $\tilde{N}_t^i$ are i.i.d. random elements of
    $D([0,\infty),\mathbb{R})$ with $\mathbb{E}[\tilde{N}_t^i]=0$ and
    $\mathbb{E}[(\tilde{N}_t^i)^{2}]<\infty$ for any $t$ (This is a
    well-known fact for Hawkes processes. See e.g. \cite{ZhuCLT}).
    Similarly, we define
    $\tilde{X}^{i}_{t}=X^{i}_{t}-m_1\int_{0}^{t}g_1(s)ds$.

    By Hahn's theorem (see e.g. Theorem 7.2.1. in \cite{whitt2002}), we
    have as $\mu \rightarrow \infty,$
    \begin{equation} \label{Hahn}
    \frac{1}{\sqrt{\mu}}\sum_{i=1}^{\mu}\left[X^{i}_{t}-m_1\int_{0}^{t}g_1(s)ds\right] \Rightarrow G,
    \end{equation}
    weakly in $(D([0,\infty),\mathbb{R}),J_{1})$, where $G$ is a
    mean-zero almost surely continuous Gaussian process with the
    covariance function of $\tilde{X}^{1}$ provided that the following
    conditions are satisfied: For every $0<T<\infty$, there exist
    continuous nondecreasing real-valued functions $g$ and $f$ on
    $[0,T]$ with numbers $\alpha>1/2$ and $\beta>1$ such that
    \begin{equation}\label{ConI}
    \mathbb{E}\left[\left(\tilde{X}_u^{1} -\tilde{X}_s^{1}
    \right)^{2}\right]\leq(g(u)-g(s))^{\alpha},
    \end{equation}
    and \begin{equation}\label{ConII}
    \mathbb{E}\left[\left(\tilde{X}_u^{1} - \tilde{X}_t^{1}\right)^{2} \left(\tilde{X}_t^{1} -\tilde{X}_s^{1}\right)^{2}\right]\leq(f(u)-f(s))^{\beta},
    \end{equation}
    for all $0\leq s\leq t\leq u\leq T$ with $u-s<1$.

    First, notice that
    \begin{align*}
    \mathbb{E}\left[\left(\tilde{X}_u^{1} -\tilde{X}_s^{1}
    \right)^{2}\right]
    &=\mathbb{E}\left[\left(\left({X}_u^{1}-{X}_s^{1}\right)-m_1\int_{s}^{u}g_1(\tau)d\tau\right)^2\right]
    \\
    &\leq 2\mathbb{E}\left[\left({X}_u^{1} -{X}_s^{1}
    \right)^{2}\right] +2m_1^{2}\left(\int_{s}^{u}g_1(\tau)d\tau\right)^2. \numberthis \label{eqn}
    \end{align*}

    By using the tower property,
    \begin{align*}
    &\mathbb{E}\left[\left({X}_u^{1} -{X}_s^{1}
    \right)^{2}\right]
    \\
    &=\mathbb{E}\left[\mathbb{E}\left[\left({X}_u^{1} -{X}_s^{1}
    \right)^{2}\Big|N^{1}\right]\right]
    \\
    &=\mathbb{E}\left[\mathbb{E}\left[\left(\sum_{i=N_s^{1}+1}^{N_u^{1}}Y_i\right)^{2}\Big|N^{1}\right]\right]
    \\
    &=\mathbb{E}\left[\left[\text{Var}\left(\sum_{i=N_s^{1}+1}^{N_u^{1}}Y_i\right)+\left(\mathbb{E}\left(\sum_{i=N_s^{1}+1}^{N_u^{1}}Y_i\right)\right)^2\right]\Big|N^{1}\right]
    \\
    &=\mathbb{E}[N_u^{1}-N_s^{1}]\text{Var}(Y_{1})
    +\mathbb{E}[(N_u^{1}-N_s^{1})^{2}](\mathbb{E}[Y_{1}])^{2}
    \\
    &=\mathbb{E}[N_u^{1}-N_s^{1}]\text{Var}(Y_{1})
    +\left[\text{Var}(N_u^{1}-N_s^{1})+(\mathbb{E}(N_u^{1}-N_s^{1}))^2\right](\mathbb{E}[Y_{1}])^{2}
    \\
    &\leq\left[\frac{1}{1-\Vert
        h\Vert_{L^{1}}}(u-s)\right]\text{Var}(Y_{1})+\left[\frac{1}{(1-\Vert
        h\Vert_{L^{1}})^3}(u-s)+\left(\frac{1}{1-\Vert
        h\Vert_{L^{1}}}(u-s)\right)^2\right](\mathbb{E}[Y_{1}])^{2}
    \\
    &\leq C_{1}(u-s). \numberthis \label{eqn}
    \end{align*}

    The first inequality in \eqref{eqn} holds because
    \begin{align*}
\mathbb{E}[N_u^{1}-N_s^{1}]
&=\int_{s}^{u}g_1(\tau)d\tau \\
&\leq g_1(\infty)(u-s)=\frac{1}{1-\Vert
	h\Vert_{L^{1}}}(u-s),
    \end{align*}

 \begin{align*}
\text{Var}[N_u^{1}-N_s^{1}]
&=\int_{s}^{u}g_2(\tau)d\tau \\
&\leq g_2(\infty)(u-s)=\frac{1}{(1-\Vert
	h\Vert_{L^{1}})^3}(u-s).
\end{align*}
    We deduce that \eqref{ConI} is satisfied with $
    g(x)=kx $ for some constant $ k $ and $ \alpha=1 $.

    Similarly, by using \eqref{eqn}, we can show that
    \begin{align*}
    &\mathbb{E}\left[\left(\tilde{X}_u^{1} - \tilde{X}_t^{1}\right)^{2} \left(\tilde{X}_t^{1} -\tilde{X}_s^{1}\right)^{2}\right]
    \\
    &=\mathbb{E}\bigg[\left(\left({X}_u^{1}-{X}_t^{1}\right)-m_1\int_{t}^{u}g_1(\tau)d\tau\right)^2
    \\
    &\qquad\qquad\cdot\left(\left({X}_t^{1}-{X}_s^{1}\right)-m_1\int_{s}^{t}g_1(\tau)d\tau\right)^2\bigg]
    \\
    &\leq\mathbb{E}\bigg[\bigg(2\left({X}_u^{1}-{X}_t^{1}\right)^2+2\left(m_1\int_{t}^{u}g_1(\tau)d\tau\right)^2\bigg)
    \\
    &\qquad\qquad\cdot\bigg(2\left({X}_t^{1}-{X}_s^{1}\right)^2+2\left(m_1\int_{s}^{t}g_1(\tau)d\tau\right)^2\bigg)\bigg]
    \\
    & \leq 4\mathbb{E}\left[\left({X}_u^{1} - X_t^{1}\right)^{2} \left({X}_t^{1} -X_s^{1}\right)^{2}\right] +C_{2}(u-s)^{2},
    \end{align*}
    and we can also compute that
    \begin{align*}
    &\mathbb{E}\left[\left({X}_u^{1} - {X}_t^{1}\right)^{2}
    \left({X}_t^{1} -{X}_s^{1}\right)^{2}\right]
    \\
    &=\mathbb{E}\Big[\left((N_u^{1}-N_t^{1})\text{Var}(Y_{1})
    +(N_u^{1}-N_t^{1})^{2}(\mathbb{E}[Y_{1}])^{2}\right)
    \\
    &\qquad\qquad\cdot\left((N_t^{1}-N_s^{1})\text{Var}(Y_{1})
    +(N_t^{1}-N_s^{1})^{2}(\mathbb{E}[Y_{1}])^{2}\right)\Big]
    \\
    &=\mathbb{E}\Big[(N_u^{1}-N_t^{1})(N_t^{1}-N_s^{1})(\text{Var}(Y_{1}))^2\Big]
    \\
    &+\mathbb{E}\Big[(N_u^{1}-N_t^{1})^2(N_t^{1}-N_s^{1})^2(\mathbb{E}[Y_{1}])^{4}\Big]
    \\
    &+\mathbb{E}\Big[(N_u^{1}-N_t^{1})^2(N_t^{1}-N_s^{1})(\mathbb{E}[Y_{1}])^{2}\text{Var}(Y_{1})\Big]
    \\
    &+\mathbb{E}\Big[(N_u^{1}-N_t^{1})(N_t^{1}-N_s^{1})^2(\mathbb{E}[Y_{1}])^{2}\text{Var}(Y_{1})
    \Big]
    \\
    &\leq
    C_{3}\mathbb{E}\Big[(N_u^{1}-N_t^{1})^2(N_t^{1}-N_s^{1})^2\Big]
    \\
    &\leq C_{4}(u-s)^{2}.
    \end{align*}

    The last two inequality holds because $ (N_u^{1}-N_t^{1})\leq
    (N_u^{1}-N_t^{1})^2 $ and $(N_t^{1}-N_s^{1})\leq
    (N_t^{1}-N_s^{1})^2  $. Also, we can conclude from the proof of
    Theorem 1 of \cite{GZ3}, $
    \mathbb{E}\Big[(N_u^{1}-N_t^{1})^2(N_t^{1}-N_s^{1})^2\Big]\leq
    C_{5}(u-s)^{2} $. Note that $ C_{i}, i\in {1,2,3,4,5} $ above are
    all constants. So we deduce that \eqref{ConII} is satisfied with $
    f(x)=\bar{k} $ for some constant $ \bar{k} $ and $\beta=2 $. Thus we
    have verified \eqref{Hahn}.

    Finally, let us identify the variance and covariance function of the Gaussian
    limit $G(t)$, for any $t\geq s$,
    \begin{align*}
    \text{Cov}(X_{t}^{1},X_{s}^{1})
    &=\mathbb{E}[X_{t}^{1}X_{s}^{1}]-\mathbb{E}[X_{t}^{1}]\mathbb{E}[X_{s}^{1}]
    \\
    &=\mathbb{E}[(X_{t}^{1}-X_{s}^{1})X_{s}^{1}]
    +\mathbb{E}[(X_{s}^{1})^{2}]-m_{1}^{2}\mathbb{E}[N_{t}^{1}]\mathbb{E}[N_{s}^{1}]
    \\
    &=m_{1}^{2}\mathbb{E}[(N_{t}^{1}-N_{s}^{1})N_{s}^{1}]
    +\mathbb{E}[(X_{s}^{1})^{2}]-m_{1}^{2}\mathbb{E}[N_{t}^{1}]\mathbb{E}[N_{s}^{1}]
    \\
    &=m_{1}^{2}\mathbb{E}[(N_{t}^{1}-N_{s}^{1})N_{s}^{1}]
    +\mathbb{E}[N_{s}^{1}](m_{2}-m_{1}^{2})+\mathbb{E}[(N_{s}^{1})^{2}]m_{1}^{2}
    -m_{1}^{2}\mathbb{E}[N_{t}^{1}]\mathbb{E}[N_{s}^{1}]
    \\
    &=m_{1}^{2}\text{Cov}(N_{t}^{1},N_{s}^{1})+\mathbb{E}[N_{s}^{1}](m_{2}-m_{1}^{2}).\numberthis
    \end{align*}

\end{proof}

\subsection{Derivation of the result in \eqref{covggsi}}
\begin{align*}
&\mbox{Cov}\left(\frac{G(t)}{\sigma(t)},\frac{G(s)}{\sigma(s)}\right)
\\
&=\frac{\mbox{Cov}(G(t),G(s))}{\sigma(t)\sigma(s)}
\\
&=\frac{m_{1}^{2}\mbox{Var}[N_s^1]+\mathbb{E}[N_s^1](m_2-m_1^2)+\mathbb{E}[N_s^{1}(N_t^{1}-N_s^{1})]-\mathbb{E}[N_s^1](\mathbb{E}[N_t^1]-\mathbb{E}[N_s^1])}{\sigma(s)\sigma(t)}
\\
&=\frac{\mbox{Var}[G(s)]+m_1^2\mathbb{E}[N_s^{1}(N_t^{1}-N_s^{1})]-\mathbb{E}[N_s^1](\mathbb{E}[N_t^1]-\mathbb{E}[N_s^1])]}{\sigma(s)^2+\sigma(s)(\sigma(t)-\sigma(s))}\\
&=\frac{\sigma(s)^2+m_1^2\mathbb{E}[N_s^1\int_{s}^{t}\lambda_{\tau}d\tau]-m_1^2\mathbb{E}[N_s^1]\mathbb{E}[\int_{s}^{t}\lambda_{\tau}d\tau]}{\sigma(s)^2+\sigma(s)(\sigma(t)-\sigma(s))}\\
&=\frac{\sigma(s)^2+m_1^2\mathbb{E}[N_s^1\lambda_s](t-s)-m_1^2\mathbb{E}[N_s^1]\mathbb{E}[\lambda_s](t-s)+o(t-s)}{\sigma(s)^2+\sigma(s)(\sigma(t)-\sigma(s))}\\
&=\frac{\sigma(s)^2+m_1^2\mbox{Cov}(N_s^1,\lambda_s)(t-s)+o(t-s)}{\sigma(s)^2+\sigma(s)(\sigma(t)-\sigma(s))}\\
&=\frac{1+\frac{m_1^2\mbox{Cov}(N_s^1,\lambda_s)}{\sigma(s)^2}(t-s)+o(t-s)}{1+\frac{\sigma(t)-\sigma(s)}{\sigma(s)}}\\
&=\left(1+\frac{m_1^2\mbox{Cov}(N_s^1,\lambda_s)}{\sigma(s)^2}(t-s)+o(t-s)\right)\left(1-\frac{\sigma(t)-\sigma(s)}{\sigma(s)}+o(t-s)\right)\\
&=1-\left(\frac{\sigma(t)-\sigma(s)}{\sigma(s)}-1\right)\frac{m_1^2\mbox{Cov}(N_s^1,\lambda_s)}{\sigma(s)^2}(t-s)-\frac{\sigma(t)-\sigma(s)}{\sigma(s)}+o(t-s)\\
&=1+\left(\frac{m_1^2(2\mbox{Cov}(N_T^1,\lambda_T)-g_2(T))-(m_2-m_1^2)g_1(T)}{2\sigma(T)^2}\right)(t-s)+o(t-s)
\end{align*}
as $t>s$, $ t-s\rightarrow 0 $ and $s\rightarrow T$.

\subsection{Derivation of the results in Remark 5 and 6}
    To compute $ \text{Cov}(N_{t}^{1},N_{s}^{1}) $, we have:
    \begin{align*}
    \text{Cov}(N_{t}^{1},N_{s}^{1})
    &=\mathbb{E}[N_s^{1}N_t^{1}]-\mathbb{E}[N_s^{1}]\mathbb{E}[N_t^{1}]\\
    &=\mathbb{E}[(N_s^{1})^2]+\mathbb{E}[N_s^{1}(N_t^{1}-N_s^{1})]-\mathbb{E}[N_s^{1}]\mathbb{E}[N_t^{1}]
    \end{align*} and
    \begin{equation} \label{expec}
\mathbb{E}[N_s^{1}(N_t^{1}-N_s^{1})]=\mathbb{E}\left[N_s^{1}\int_{s}^{t}\lambda_{\tau}d\tau\right].
    \end{equation}

    Here $ \lambda_{\tau}=1+\int_{0}^{\tau}h(\tau-s)dN_s^{1} =1+Z_{\tau}$,
    so \eqref{expec} can be written as:

    \begin{align*}
    	\mathbb{E}[N_s^{1}(N_t^{1}-N_s^{1})] & =\mathbb{E}\left[N_s^{1}\int_{s}^{t}(1+Z_{\tau})d\tau\right]\\
    	&=(t-s)\mathbb{E}[N_s^{1}]+\int_{s}^{t}\mathbb{E}[Z_{\tau}N_s^{1}]d\tau.\numberthis\label{expect1}
    \end{align*}

    In the special case when $ h(t)=\alpha e^{-\beta t} $, $ Z_{\tau} $ is Markovian and we can get the explicit formula.

   We have $ dZ=-\beta Zdt+\alpha dN_t^{1} $, so

   \begin{equation*}
   N_t^{1}=\frac{Z_t-Z_0}{\alpha}+\frac{\beta}{\alpha}\int_{0}^{t}Z_udu.
   \end{equation*}

   In \eqref{expect1},

   \begin{align*}
\int_{s}^{t}\mathbb{E}[Z_{\tau}N_s^{1}]d\tau
&=\int_{s}^{t}\mathbb{E}\left[Z_{\tau}\left(\frac{Z_s-Z_0}{\alpha}\right)+\frac{Z_{\tau}\beta}{\alpha}\int_{0}^{s}Z_udu\right]d\tau\\
&=\int_{s}^{t}\left[\frac{1}{\alpha}\mathbb{E}[Z_{\tau}Z_s]-\frac{Z_0}{\alpha}\mathbb{E}[Z_{\tau}]+\mathbb{E}\left[\frac{Z_{\tau}\beta}{\alpha}\int_{0}^{s}Z_udu\right]\right]d\tau\\
&=\int_{s}^{t}\left[\frac{1}{\alpha}\mathbb{E}[Z_{\tau}Z_s]-\frac{Z_0}{\alpha}\mathbb{E}[Z_{\tau}]+\frac{\beta}{\alpha}\int_{0}^{s}\mathbb{E}\left[Z_{\tau}Z_u\right]du\right]d\tau.\numberthis\label{expectzn}
   \end{align*}

    Here $ \mathbb{E}[Z_{\tau}Z_s]=\mathbb{E}\left[\mathbb{E}[Z_{\tau}|Z_s]Z_s\right] $, $ s<\tau $. Also we have:

    \begin{equation*}
    d\mathbb{E}[Z_{\tau}]=-\beta\mathbb{E}[Z_{\tau}]d{\tau}+\alpha(1+\mathbb{E}[Z_{\tau}])d{\tau}
    \end{equation*}
and
    \begin{equation*}
\frac{d\mathbb{E}[Z_{\tau}]}{d\tau}=(\alpha-\beta)\mathbb{E}[Z_{\tau}]+\alpha.
    \end{equation*}

    Solving it, we get:

    \begin{equation}\label{expectz}
    \mathbb{E}[Z_{\tau}]=\frac{\alpha}{\alpha-\beta}e^{(\alpha-\beta)\tau}-\frac{\alpha}{\alpha-\beta}.
    \end{equation}
    Moreover, we have
    \begin{align*}
    e^{(\beta-\alpha)\tau}\mathbb{E}[Z_{\tau}|Z_s]
    &=e^{(\beta-\alpha)s}Z_s+\int_{s}^{\tau}\alpha e^{(\beta-\alpha)u}du\\
    &=e^{(\beta-\alpha)s}Z_s+\frac{\alpha}{\beta-\alpha}\left[e^{(\beta-\alpha)\tau}-e^{(\beta-\alpha)s}\right]
    \end{align*}
and
    \begin{equation*}
\mathbb{E}[Z_{\tau}|Z_s]=e^{(\beta-\alpha)(s-\tau)}Z_s+\frac{\alpha}{\beta-\alpha}\left[1-e^{(\beta-\alpha)(s-\tau)}\right].
    \end{equation*}

    Hence we have:
    \begin{equation}\label{expectzz}
    \mathbb{E}[Z_{\tau}Z_s]=e^{(\beta-\alpha)(s-\tau)}\mathbb{E}[Z_s^2]+\frac{\alpha}{\beta-\alpha}\left[1-e^{(\beta-\alpha)(s-\tau)}\right]\mathbb{E}[Z_s].
    \end{equation}

    Next, we need to figure out $ \mathbb{E}[Z_s^2] $ by using the infinitesimal generator as following:
    \begin{equation}
    \mathcal{A}f(z)=-\beta zf'(z)+(1+z)[f(z+\alpha)-f(z)],
    \end{equation}
Let $f(z)=z^2,$ then we have 
    \begin{align*}
    \mathcal{A}(z^2)
    &=-\beta z(z^2)'+(1+z)\left[(z+\alpha)^2-z^2\right]\\
    &=-2\beta z^2+(1+z)(2\alpha z+\alpha^2)\\
    &=2(\alpha-\beta)z^2+(2\alpha+\alpha^2)z+\alpha^2,
    \end{align*}
and hence, we have
    \begin{align*}
    \mathbb{E}[Z_{\tau}^2]
    &=Z_0^2+\int_{0}^{\tau}\mathbb{E}[\mathcal{A}(Z_s^2)]ds\\
    &=Z_0^2+\int_{0}^{\tau}\left[2(\alpha-\beta)\mathbb{E}[Z_s^2]+(2\alpha+\alpha^2)\mathbb{E}[Z_s]+\alpha^2\right].
    \end{align*}

    By differentiation, we get:
    \begin{equation*}
    \frac{d\mathbb{E}[Z_{\tau}^2]}{d\tau}=2(\alpha-\beta)\mathbb{E}[Z_{\tau}^2]+(2\alpha+\alpha^2)\mathbb{E}[Z_{\tau}]+\alpha^2.
    \end{equation*}

    Solving it, we get

    \begin{equation}\label{expectz2}
\mathbb{E}[Z_{\tau}^2]=\frac{-\alpha(2\alpha+\alpha^2)e^{(\beta-\alpha)\tau}}{(\beta-\alpha)^2}+\frac{\alpha^2(2+\beta)}{2(\beta-\alpha)^2}+\frac{\alpha^2(2+2\alpha-\beta)}{2(\beta-\alpha)^2}e^{2(\alpha-\beta)\tau}.
    \end{equation}
    
    By substituting \eqref{expectz} and \eqref{expectz2} back into \eqref{expectzz} we get:
    \begin{equation}
    \begin{split}
\mathbb{E}[Z_{\tau}Z_s]&=\frac{-\alpha(2\alpha+\alpha^2)}{(\beta-\alpha)^2}e^{[(\beta-\alpha)s+2(\alpha-\beta)\tau]}+\frac{\alpha^2(2+\beta)}{2(\beta-\alpha)^2}e^{(\beta-\alpha)(s+\tau)}\\
&+\frac{\alpha^2(2+2\alpha-\beta)}{2(\beta-\alpha)^2}e^{(\beta-\alpha)(s-\tau)}-\frac{\alpha^2}{(\beta-\alpha)^2}e^{(\alpha-\beta)\tau}+\frac{\alpha^2}{(\beta-\alpha)^2}e^{(\beta-\alpha)(s-2\tau)}\\
&-\frac{\alpha}{(\beta-\alpha)^2}e^{(\beta-\alpha)(s-\tau)}+\frac{\alpha}{(\beta-\alpha)^2}.
    \end{split}
    \end{equation}

    Substitute into \eqref{expectzn}, we get:
    \begin{equation}
    \begin{split}
&\mathbb{E}[Z_{\tau}N_s^{1}]=\frac{1}{\alpha}\mathbb{E}[Z_{\tau}Z_s]-\frac{Z_0}{\alpha}\mathbb{E}[Z_{\tau}]+\frac{\beta}{\alpha}\int_{0}^{s}\mathbb{E}\left[Z_{\tau}Z_u\right]du\\
&=M_1e^{(\beta-\alpha)s}+M_2e^{(\beta-\alpha)\tau}+M_3e^{(\alpha-\beta)\tau}+M_4e^{2(\alpha-\beta)\tau},
    \end{split}
    \end{equation}

    where $ M_i, i\in 1,2,3,4 $ are constants and

    \begin{equation*}
    \begin{split}
 &M_1=\frac{(2\beta-\alpha)[-2\alpha^2(2+\alpha)+\alpha^2(2+\beta)+\alpha^2(2+2\alpha-\beta)-2\alpha]}{2\alpha(\beta-\alpha)^3}\\
 &M_2=\frac{\alpha(2+\beta)}{2(\beta-\alpha)^2}\\
 &M_3=\frac{\alpha(2+2\alpha-\beta)(2\beta-\alpha-\alpha\beta)-(\beta-\alpha)(2\alpha+2)}{2(\beta-\alpha)^3}\\
 &M_4=\frac{-\alpha(3\beta+\alpha\beta-2\alpha-\alpha^2)}{(\beta-\alpha)^3}.
    \end{split}
    \end{equation*}

    Hence, we have:

    \begin{equation}
    \begin{split}
 &\int_{s}^{t}\mathbb{E}[Z_{\tau}N_s^{1}]d\tau=M_1e^{(\alpha-\beta)s}(t-s)+M_2\left[e^{(\beta-\alpha)t}-e^{(\beta-\alpha)s}\right]\\
 &+M_3\left[e^{(\alpha-\beta)t}-e^{(\alpha-\beta)s}\right]+M_4\left[e^{2(\alpha-\beta)t}-e^{2(\alpha-\beta)s}\right].
    \end{split}
    \end{equation}

    According to \eqref{expect1}, we have

    \begin{equation}\label{expectofnn}
    \begin{split}
	&\mathbb{E}[N_s^{1}(N_t^{1}-N_s^{1})]=\left[M_1e^{(\alpha-\beta)s}+\frac{\alpha}{(\alpha-\beta)^2}\left[e^{(\alpha-\beta)s}-1\right]-\frac{\beta s}{\alpha-\beta}\right](t-s)\\
	&+M_2\left[e^{(\beta-\alpha)t}-e^{(\beta-\alpha)s}\right]+M_3\left[e^{(\alpha-\beta)t}-e^{(\alpha-\beta)s}\right]+M_4\left[e^{2(\alpha-\beta)t}-e^{2(\alpha-\beta)s}\right].
    \end{split}
    \end{equation}

    Thus, we can compute that $\mbox{Cov}(G(t),G(s))$ as following:
    \begin{align*}
    &\mbox{Cov}(G(t),G(s))
    \\
    &=\mbox{Cov}(X_t^1,X_s^1)
    \\
    &=m_1^2\left[\mathbb{E}[(N_s^{1})^2]+\mathbb{E}[N_s^{1}(N_t^{1}-N_s^{1})]-\mathbb{E}[N_s^{1}]\mathbb{E}[N_t^{1}]\right]+\mathbb{E}[N_s^1](m_2-m_1^2)\\
    &=m_1^2\bigg(\mbox{Var}[N_s^1]+(\mathbb{E}[N_s^1])^2+\mathbb{E}[N_s^{1}(N_t^{1}-N_s^{1})]-\mathbb{E}[N_s^{1}]\mathbb{E}[N_t^{1}]\bigg)+\mathbb{E}[N_s^1](m_2-m_1^2).\\   \numberthis \label{covgg}
    \end{align*}

    By substituting \eqref{expectofn}, \eqref{varofn} and \eqref{expectofnn} into \eqref{covgg}, we finally obtain the covariance of $ G $.

    \begin{align*}
    &\mbox{Cov}(G(t),G(s))\\
    &=m_1^2\bigg(\frac{\alpha^2}{(\alpha-\beta)^3}\left[e^{2(\alpha-\beta)t}-1\right]-\frac{\alpha}{(\alpha-\beta)^2}\left[e^{(\alpha-\beta)t}-1\right]-\frac{2\alpha\beta t}{(\alpha-\beta)^2}e^{(\alpha-\beta)t}+\frac{\beta t}{\beta-\alpha}\\
    &+\left[\frac{\alpha}{(\alpha-\beta)^2}\left[e^{(\alpha-\beta)s}-1\right]-\frac{\beta s}{\alpha-\beta}\right]^2\\
    &+\left[M_1e^{(\alpha-\beta)s}+\frac{\alpha}{(\alpha-\beta)^2}\left[e^{(\alpha-\beta)s}-1\right]-\frac{\beta s}{\alpha-\beta}\right](t-s)\\
    &+M_2\left[e^{(\beta-\alpha)t}-e^{(\beta-\alpha)s}\right]+M_3\left[e^{(\alpha-\beta)t}-e^{(\alpha-\beta)s}\right]+M_4\left[e^{2(\alpha-\beta)t}-e^{2(\alpha-\beta)s}\right]\\
    &-\left[\frac{\alpha}{(\alpha-\beta)^2}\left[e^{(\alpha-\beta)s}-1\right]-\frac{\beta s}{\alpha-\beta}\right]\left[\frac{\alpha}{(\alpha-\beta)^2}\left[e^{(\alpha-\beta)t}-1\right]-\frac{\beta t}{\alpha-\beta}\right]\bigg)\\
    &+(m_2-m_1^2)\bigg(\frac{\alpha}{(\alpha-\beta)^2}\left[e^{(\alpha-\beta)s}-1\right]-\frac{\beta s}{\alpha-\beta}\bigg). \numberthis
    \end{align*}

In \eqref{covgg}, setting $ s=t $, we obtain the variance of $ G $:

\begin{align*}
&\mbox{Var}(G(t))\\
&=m_1^2\bigg(\frac{\alpha^2}{(\alpha-\beta)^3}\left[e^{2(\alpha-\beta)t}-1\right]-\frac{\alpha}{(\alpha-\beta)^2}\left[e^{(\alpha-\beta)t}-1\right]-\frac{2\alpha\beta t}{(\alpha-\beta)^2}e^{(\alpha-\beta)t}+\frac{\beta t}{\beta-\alpha}\bigg)\\
&+(m_2-m_1^2)\bigg(\frac{\alpha}{(\alpha-\beta)^2}\left[e^{(\alpha-\beta)t}-1\right]-\frac{\beta t}{\alpha-\beta}\bigg). \numberthis
\end{align*}

{\add \noindent
	{\bf Acknowledgements}\\
	Youngsoo Seol is grateful to the support from the Dong-A University
	research grant.\\}


\end{document}